\documentstyle[11pt,newpasp,twoside]{article}
\def\etal{{et\thinspace al.}\ }
\begin{document}
\title{Model Photospheres with Accelerated Lambda Iteration}
\author{Klaus Werner, Jochen L. Deetjen, Stefan Dreizler, Thorsten Nagel,
        Thomas Rauch, Sonja L. Schuh}
\affil{Institut f\"ur Astronomie und Astrophysik, Universit\"at T\"ubingen, 
Sand 1, D-72076 T\"ubingen, Germany}

\begin{abstract}
We address the classical stellar-atmosphere problem and describe our method of
numerical solution in detail. The problem consists of the solution of the
radiation transfer equation under the constraints of hydrostatic, radiative and
statistical equilibrium (non-LTE). We employ the Accelerated Lambda Iteration
(ALI) technique, and use statistical methods to construct non-LTE 
metal-line-blanketed model-atmospheres.
\end{abstract}

\section{Introduction}

Essentially all our knowledge about the structure and evolution of stars, hence
about galactic evolution in general, rests on the interpretation of their
electromagnetic spectrum. Therefore quantitative analysis of stellar spectra is
one of the most important tools of modern astrophysics. The formation of the
observed spectrum is confined to the atmosphere, a very thin layer on top of
the stellar core. Spectral analysis is performed by modeling the
temperature- and pressure-stratification of the atmosphere and computing
synthetic spectra which are then compared to observation. Fitting synthetic
spectra to the data yields basic photospheric parameters: effective temperature,
surface gravity, and chemical composition. Comparison with theoretical
evolutionary calculations allows one to derive stellar parameters: mass,
radius, and total luminosity.

\mbox{\qquad}The classical stellar-atmosphere problem considers radiation transfer through
the outermost layers of a star into free space under three assumptions. First
it is assumed that the atmosphere is in hydrostatic equilibrium, thus, the
matter which interacts with photons is at rest. Second, the transfer of energy
through the atmosphere is assumed to be done entirely by photons, i.e.\
heat-conduction and convection are regarded as negligible (radiative
equilibrium). But the effectiveness of photon-transfer depends on the opacity
and emissivity of the matter, which are strongly state- and frequency-dependent
quantities. They depend in detail on the occupation numbers of atomic/ionic
levels, which, in turn, are determined by the local temperature and electron
density, and the radiation field, whose nature is {\it non-local}. The
occupation number of any atomic level is determined by a balance among
radiative and collisional population and de-population processes (statistical
equilibrium; our third assumption), i.e.\ the interaction of atoms with other
particles and photons. Mathematically, the whole problem requires the solution
of the radiation transfer equations simultaneously with the equations for
hydrostatic and radiative equilibrium, together with the statistical
equilibrium (rate equations).

A stellar atmosphere radiates into circumstellar space and thus is an {\it open
thermodynamic system}; hence it cannot be in thermodynamic equilibrium (TE),
and cannot be characterized by a single temperature. The idea of ``Local
Thermodynamic Equilibrium'' (LTE) is a hypothesis which assumes that while TE
does not hold for the atmosphere as a whole, it can be applied in small volume
elements. In this case the atomic occupation numbers depend only on the local
electron temperature and electron density via the Saha-Boltzmann equations.
But this approximation can be valid only in the limit that collision rates
dominate radiative rates, and photon-mean-free-paths are small. Models in which
the Saha-Boltzmann equations are replaced by the physically more accurate rate
equations are called non-LTE (or NLTE) models. This designation is somewhat
imprecise because the velocity distribution of particles is still assumed to be
Maxwellian, i.e.\ we can still define a local temperature. NLTE calculations
are more costly than LTE calculations, however, and it is hard to predict a
priori whether NLTE effects will be important in a specific problem. Generally,
NLTE effects are large at high temperatures and low densities, which imply
intense radiation fields, hence frequent radiative processes (which have
a non-local nature and, in particular, respond to the presence of the open
boundary), and less-frequent particle collisions, which tend to enforce LTE
conditions.

Abandonment of the LTE assumption leads to a much more difficult
model-atmosphere problem, because of subtle couplings between the radiation
transfer and statistical equilibrium equations. The pioneering work by Auer
\& Mihalas (1969) provided a basic tool for making such models. But the
numerical problem going from LTE to {\it realistic} NLTE models has been solved
only recently, and is the topic of this paper. We now have much more powerful
tools to compute non-classical models, which account for very complex opacities,
and solve the radiation transfer equations in more general environments, for
example in expanding stellar atmospheres. These are the topics of other papers
in this volume.

Stellar-atmosphere modeling has made huge progress in recent years. This
advance was achieved by the development of new numerical techniques for
model-construction, and by the availability of atomic data for many species.
And these achievements became possible only with an enormous increase of
computing power. Model atmospheres assuming LTE have been highly refined by the
inclusion of many more atomic and molecular opacity sources, and effective
numerical techniques for LTE model-computation have been available for years.
The progress is most remarkable in the field of NLTE model atmospheres.
Replacement of the Saha-Boltzmann equations by atomic rate-equations requires
radically different numerical-solution techniques, otherwise metal-opacities
cannot be taken into account at all. Such techniques have been developed with
great success during the last decade, inspired by important papers by Cannon
(1973) and Scharmer (1981). The {\it Accelerated Lambda-Iteration Method}
(ALI) is at the heart of this development. Combined with statistical
representations of line-opacities, we are finally able to compute 
metal-line-blanketed NLTE models, including many millions of spectral lines,
with a very high level of sophistication.

In this paper we discuss the basic ideas behind the new numerical methods for
NLTE modeling. We begin by presenting the classical model-atmosphere problem
and introduce the basic equations. Then we focus on the ALI solution-method,
and our numerical implementation of it. We then briefly describe the solution
of the NLTE metal-line-blanketing problem.

\section{Statement of the problem and overview of the solution
method\label{eins2}}

We assume plane-parallel geometry, which is well justified for most stars
because the atmospheres are thin compared to the stellar radius. The only
parameters needed to characterize uniquely such an atmosphere are the effective
temperature (\hbox{$T_{\rm eff}$}), a measure for the amount of energy
transported through the atmosphere per unit area and time (see
Eq.\,\ref{nominal}), the surface gravity ($g$), and the chemical composition.
Generalization to spherical symmetry to account for extended (static)
atmospheres mainly affects the radiation transfer equation and is
straightforward (Mihalas \& Hummer 1974). A spherical-symmetry variant of our
code has been written by Nagel \etal (2001). To construct model atmospheres we
must solve simultaneously a set of equations that is highly coupled and
non-linear. Because of the coupling, no equation determines uniquely a single
quantity -- {\it all} equations jointly determine each of the state parameters.
Neverlethess, each of them is usually thought of as determining, more or less,
a particular quantity. These equations are:

\begin{itemize}
\item The radiation transfer equations which are solved for the angle-integrated mean
intensities $J_i, i=1,\ldots,NF$, on a pre-chosen frequency grid comprising
$NF$ points. The formal solution is given by $J=\Lambda S$, where $S$ is the
source function as defined later (Eq.\,\ref{source}), and $\Lambda$ is the
transport operator. Although $\Lambda$ is written as an operator, one may think
of it as a {\em process} of obtaining the mean intensity from the source
function.
\item The hydrostatic-equilibrium equation which determines the total particle
density $N$.
\item The radiative-equilibrium equation from which the temperature $T$ follows.
\item The particle-conservation equation, determining the electron density 
$n_e$.
\item The statistical-equilibrium equations, which are solved for the population
densities $n_i, i=1,\ldots,NL$  of the atomic levels that are allowed to depart
from LTE (NLTE levels).
\item The equation that defines a fictitious massive-particle density $n_H$,
which is introduced for convenience in the solution procedure.
\end{itemize}

This set of equations has to be solved at each point $d$ of a grid comprising
$ND$ depth points. Thus we are looking for solution vectors
\begin{equation}\label{psi1}
\mbox{\boldmath$\psi$}_d' = (n_1,\ldots,n_{NL}, n_e, T, n_H, N,
J_1,\ldots,J_{NF}) , \quad d=1, \ldots ,ND .
\end{equation}
The Complete Linearization (CL) method (Auer \& Mihalas 1969) solves this set by
linearizing the equations with respect to all variables. The basic advantage of
the ALI (or ``operator splitting'') method over CL is that it allows one to
eliminate, at the outset, the explicit appearance of the mean intensities
$J_i$ from the solution scheme by expressing these variables by the current,
yet-to-be determined, occupation numbers and temperature. This is accomplished
by an iteration procedure which may be written as (suppressing indices
indicating depth and frequency dependence of variables):
\begin{equation}\label{ali}
J^n=\Lambda^{\star}S^n+(\Lambda-\Lambda^{\star})S^{n-1} .
\end{equation}
This means that the actual mean intensity at any iteration step $n$ is computed
by applying an approximate lambda operator (ALO) $\Lambda^{\star}$ on the
actual source function $S^n$ plus a correction term that is computed
from quantities known from the previous iteration step. This correction term
includes the exact lambda operator $\Lambda$, which guarantees convergence to
the exact solution of the radiation transfer problem
$J=\Lambda S$. The use of $\Lambda$ in Eq.\,\ref{ali} only indicates that a
formal solution of the transfer equation is performed, but in our application
the operator is not constructed explicitly. Instead we employ the Feautrier
(1964) solution scheme (Mihalas 1978) or a short-characteristic method (Olson
\& Kunasz 1987) to solve the transfer equation as a differential
equation. The resulting set of equations for the reduced solution vectors
\begin{equation}\label{psi2}
\mbox{\boldmath$\psi$}_d = (n_1,\ldots,n_{NL}, n_e, T, n_H, N) , \quad
d=1,\ldots,ND
\end{equation}
is of course still non-linear. The solution is obtained by linearization and
iteration which is performed either with a usual Newton-Raphson iteration or by
other, much faster methods like the quasi-Newton or Kantorovich variants
(Dreizler \& Werner 1991, Hubeny \& Lanz 1992) (see Sect.\,\ref{broy}). The
first model-atmosphere calculations with the ALI method were performed by
Werner (1986).

Another advantage of the ALI method is that explicit depth coupling of the
solution vectors Eq.\,\ref{psi1} through the transfer equation can be avoided
if one uses diagonal (i.e.\ local) ALOs. Then the solution vectors
Eq.\,\ref{psi2} are independent from each other and the solution procedure
within one iteration step of Eq.\,\ref{ali} is much more straightforward. Depth
coupling is provided by the correction which contains the exact solution of
the transfer equation. The hydrostatic equation, which also gives explicit
depth coupling, may be taken out of the set of equations and can -- as
experience shows -- be solved in between two iteration steps of Eq.\,\ref{ali}.
Then full advantage of a local ALO can be taken. The linearized system may be
written as:
\begin{equation}
\mbox{\boldmath$\psi$}_d
=\mbox{\boldmath$\psi$}_d^0+\delta\mbox{\boldmath$\psi$}_d \ ,
\end{equation}
where $\mbox{\boldmath$\psi$}_d^0$ is the current estimate for the solution
vector at depth $d$ and $\delta\mbox{\boldmath$\psi$}_d$ is the correction
vector to be computed. Using a tri-diagonal $\Lambda^{\star}$ operator the
resulting system for $\delta\mbox{\boldmath$\psi$}_d$ is -- like in the
classical CL scheme -- of block tri-diagonal form coupling each depth point $d$
to its nearest neighbors $d\pm1$:
\begin{equation}\label{tri}
\mbox{\boldmath$\gamma$}_d\delta\mbox{\boldmath$\psi$}_{d-1}+
\mbox{\boldmath$\beta$}_d\delta\mbox{\boldmath$\psi$}_d+
\mbox{\boldmath$\alpha$}_d\delta\mbox{\boldmath$\psi$}_{d+1}={\bf c}_d .
\end{equation}
The quantities {$\mbox{\boldmath$\alpha, \beta, \gamma$}$} are ($NN\times NN$)
matrices where $NN$ is the total number of physical variables, i.e., $NN=NL+4$,
and ${\bf c}_d$ is the residual error in the equations. The solution is
obtained by the standard Feautrier scheme. With starting values ${\bf
D}_1=\mbox{\boldmath$\beta$}_1^{-1}(\mbox{-\boldmath$\alpha$}_1$) and ${\bf
v}_1=\mbox{\boldmath$\beta$}_1^{-1}{\bf c}_1$ we sweep from the outer boundary
of the atmosphere inside and calculate at each depth:
\begin{eqnarray}\label{feau}
{\bf D}_d&=&(\mbox{\boldmath$\beta$}_d+
\mbox{\boldmath$\gamma$}_d{\bf D}_{d-1})^{-1}
(-\mbox{\boldmath$\alpha$}_d)\nonumber\\
{\bf v}_d&=&
(\mbox{\boldmath$\beta$}_d+\mbox{\boldmath$\gamma$}_d{\bf D}_{d-1})^{-1}
({\bf c}_d-\mbox{\boldmath$\gamma$}_d{\bf v}_{d-1}) .
\end{eqnarray}
At the inner boundary we have $\mbox{$\bf D$}_{ND}=\mbox{\bf 0}$ and by
sweeping backwards to the surface we calculate the correction vectors, first
$\delta\mbox{\boldmath$\psi$}_{ND}={\bf v}_{ND}$, and then successively
$\delta\mbox{\boldmath$\psi$}_d={\bf
D}_d\delta\mbox{\boldmath$\psi$}_{d+1}+{\bf v}_d$. As already mentioned, the
system Eq.\,\ref{tri} breaks into $ND$ independent equations
$\delta\mbox{\boldmath$\psi$}_d=\mbox{\boldmath$\beta$}_d^{-1}{\bf c}_d$
($d=1,\ldots,ND$) when a local $\Lambda^{\star}$ operator is used. The
additional numerical effort to set up the subdiagonal matrices and matrix
multiplications in the tri-diagonal case is outweighed by the faster global
convergence of the ALI cycle, accomplished by the explicit depth coupling in the
linearization procedure (Werner 1989).

The principal advantage of the ALI over the CL method becomes clear at this
point. Each matrix inversion in Eq.\,\ref{feau} requires $(NL+4)^3$ operations
whereas in the CL method $(NL+NF+4)^3$ operations are needed. Since the number
of frequency points $NF$ is much larger than the number of levels $NL$, the
matrix inversion in the CL approach is dominated by $NF$.

Recent developments deal with the problem that the total number of atomic levels
tractable in NLTE with the ALI method described so far is restricted to the
order of 250, from experience with our model atmosphere code {\tt PRO2}. This
limit is a consequence of the non-linearity of the equations, and in order to
overcome it,  measures must be taken in order to achieve a linear system whose
numerical solution is much more stable. Such a pre-conditioning procedure was
first applied in the ALI context by Werner \& Husfeld (1985) using the
core-saturation method (Rybicki 1972). More advanced work achieves linearity by
replacing the $\Lambda$ operator with the $\Psi$ operator (and by judiciously
considering some populations as ``old'' and some as ``new'' ones within an ALI
step) which is formally defined by writing:
\begin{equation}
J_\nu=\Psi_\nu\eta_\nu\ , \qquad {\rm i.e.} \qquad
\Psi_\nu\equiv\Lambda_\nu/\chi_\nu\ ,
\end{equation}
where the total opacity $\chi_\nu$ (as defined in Sect.\,\ref{opa}) is
calculated from the previous ALI cycle. The advantage is that the emissivity
$\eta_\nu$ (Sect.\,\ref{opa}) is linear in the populations, whereas the source
function $S_\nu$ is not. Hence the new operator $\Psi$ gives the solution of the
transfer problem by acting on a linear function. This idea was proposed by 
Rybicki \& Hummer (1991) who applied it to the line-formation problem, i.e.\
regarding the atmospheric structure as fixed. Hauschildt (1993) and Hauschildt
\& Baron (1999) generalized it to solve the full model-atmosphere problem. In
addition, splitting the set of statistical equations, and solving it separately
for each chemical element, means that now many hundreds of levels per species
are tractable in NLTE (see Dreizler, this volume, for our 
experience). A very robust method and fast variant of the ALI method, the
ALI/CL hybrid scheme, permits the linearization of the radiation field for
selected frequencies (Hubeny \& Lanz 1995), but it is not implemented in {\tt
PRO2}.

\section{Basic equations}

\subsection{Radiation transfer}

Any numerical method requires a formal solution (i.e.\ atmospheric structure
already given) of the radiation transfer problem. We disregard here
polarization effects. (An implementation of an ALI technique to treat
polarized radiation is discussed by Deetjen \etal in this
volume.) Radiation transfer at any depth point is described by the transfer
equation:
\begin{equation}\label{te}
\pm\mu \frac{\partial I_{\nu}(\pm\mu)}{\partial\tau_{\nu}}=
S_{\nu}-I_{\nu}(\pm\mu), \quad\mu\in[0,1],
\end{equation}
written separately for positive and negative $\mu$ (the angle-cosine between
the direction of propagation and the outward-directed normal to the surface),
i.e.\ for the inward and outward directional intensities $I(\mu)$ at frequency
$\nu$. Here $\tau_\nu$ is the optical depth (defined in terms of the column
mass $m$ used in the other structural equations by $d\tau_\nu=dm \chi_\nu/\rho$,
where $\rho$ is the mass density), and $S_{\nu}$ is the local source function.
Introducing the Feautrier intensity
\begin{equation}\label{udef}
u_{\nu\mu}\equiv \left(I_{\nu}(\mu)+I_{\nu}(-\mu)\right)/2
\end{equation}
we obtain the second-order form (Mihalas 1978, p.\,151):
\begin{equation}
\mu^2 \frac{\partial^2u_{\nu\mu}}{\partial\tau_{\nu}^2}=u_{\nu\mu}-S_{\nu}, \quad\mu\in[0,1].
\end{equation}
We may separate the Thomson emissivity term (scattering from free electrons,
assumed coherent, with cross-section $\sigma_e$) from the source function so
that
\begin{equation}\label{source}
S_{\nu}=S_{\nu}'+n_e\sigma_eJ_{\nu}/\chi_{\nu},
\end{equation}
where $S_{\nu}'$ is the ratio of thermal emissivity to total opacity as
described in detail below (Sect.\,\ref{opa}): $S_{\nu}'=\eta_{\nu}/\chi_{\nu}$.
The mean intensity is the angular integral over the Feautrier intensity, hence
the transfer equation becomes
\begin{equation}\label{unm}
\mu^2 \frac{\partial^2u_{\nu\mu}}{\partial\tau_{\nu}^2}
=u_{\nu\mu}-S_{\nu}'-\frac{n_e\sigma_e}{\chi_{\nu}}\int_{0}^{1}u_{\nu\mu}\,d\mu .
\end{equation}
Thomson scattering complicates the situation by its explicit angle coupling but
the solution can be obtained with the standard Feautrier scheme. Assuming
complete frequency-redistribution in spectral lines (Mihalas 1978, p.\,29), no
explicit frequency coupling occurs so that a parallel solution for all
frequencies enables very efficient vectorization on the computer.

The following boundary conditions are used for the transfer equation. At the
inner boundary where the optical depth is at maximum, $\tau=\tau_{\rm max}$, we
have
\begin{equation}
\left(\mu \frac{\partial u_{\nu\mu}}{\partial\tau_{\nu}}\right)_{\tau_{\rm max}}
=I^+_{\nu\mu}-u_{\nu\mu}(\tau_{\rm max}),
\end{equation}
where we specify $I^+_{\nu\mu}=I_{\nu}(+\mu, \tau_{\rm max})$ from the
diffusion approximation:
\begin{equation}\label{inner}
I^+_{\nu\mu}=B_{\nu}+\frac{3\mu}{\chi_{\nu}} \frac{\partial B_{\nu}}{\partial T} 
\frac{{\cal H}}
{\int_{0}^{\infty}\frac{1}{\chi_\nu}\frac{\partial B_{\nu}}{\partial T}\,d\nu}.
\end{equation}
Here $B_{\nu}$ is the Planck function and ${\cal H}$ the nominal (frequency-integrated) Eddington flux:
\begin{equation}\label{nominal}
{\cal H}=\sigma_R T_{\rm eff}^4/4\pi,
\end{equation}
with the Stefan-Boltzmann constant $\sigma_R$. At the outer boundary we take
$\tau_\nu=\tau_{\rm min}=m_1 \chi_\nu/2\rho$, assuming that $\chi$ is a linear
function of $m$ for $m<m_1$. Since $\tau_{\rm min}\neq 0$, it is not exactly
valid to assume no incident radiation at the stellar surface. Instead we
specify $I^-_{\nu\mu}=I_{\nu}(-\mu, \tau_{\rm min})$ after Scharmer \& Nordlund
(1982):
\begin{equation}
I^-_{\nu\mu}=S_\nu(\tau_{\rm min})[1-\exp(-\tau_{\rm min}/\mu)]
\end{equation}
which follows from Eq.\,\ref{te} assuming $S(\tau)=S(\tau_{\rm min})$ for
$\tau<\tau_{\rm min}$. Then we get:
\begin{equation}
\left(\mu \frac{\partial u_{\nu\mu}}{\partial\tau_{\nu}}\right)_{\tau_{\rm min}}
=u_{\nu\mu}(\tau_{\rm min})-I^-_{\nu\mu}.
\end{equation}
The boundary conditions are discretized performing Taylor expansions which
yield second-order accuracy (Mihalas 1978, p.\,155).

\subsection{Statistical equilibrium}

The statistical equilibrium equations are set up according to Mihalas (1978,
p.\,127). The number of atomic levels, ionization stages, and chemical species,
as well as all radiative and collisional transitions are taken from the input
model atom supplied by the user. Ionization into excited states of
the next ionization stage is taken into account. Dielectronic recombination and
autoionization processes can also be included in the model atom.

\subsubsection{Rate equations}

As usual, the atomic energy levels are ordered sequentially by increasing
excitation energy, starting with the lowest ionization stage. Then for each
atomic level $i$ of any ionization stage of any element, the rate equation
describes the equilibrium of rates into and rates out of this level:
\begin{equation}\label{rates}
n_i\sum_{i\neq j}^{}P_{ij}-\sum_{j\neq i}^{}n_jP_{ji}=0 .
\end{equation}
The rate coefficients $P_{ij}$ have radiative and collisional components:
$P_{ij}=R_{ij}+C_{ij}$. Radiative upward and downward rates are respectively
given by:
\begin{equation}
R_{ij}=4\pi\int_{0}^{\infty} \frac{\sigma_{ij}(\nu)}{h\nu}J_{\nu}\,d\nu,
\end{equation}
\begin{equation}
R_{ji}=\left(\frac{n_i}{n_j}\right)^{\star}4\pi\int_{0}^{\infty} 
\frac{\sigma_{ij}(\nu)}
{h\nu}\left(\frac{2h\nu^3}{c^2}+J_{\nu}\right)e^{-h\nu/kT}\,d\nu .
\end{equation}
Photon cross-sections are denoted by $\sigma_{ij}(\nu)$.
$({n_i}/{n_j})^{\star}$ is the Boltzmann LTE population ratio in the case of
line transitions: $g_i/g_j \exp (-h\nu_{ij}/kT)$, where the $g_{i,j}$ are the
statistical weights. The LTE population number of a particular level is defined
relative to the ground state of the next ion, so that in the case of
recombination from a ground state $n_1^+$ we have by definition
$({n_i}/{n_j})^{\star}=n_e\phi_i(T)$ with the Saha-Boltzmann factor
\begin{equation}
\phi_i(T)=2.07 \cdot 10^{-16}\frac{g_i}{g_1^+}T^{-3/2}e^{h\nu_i/kT},
\end{equation}
where $h\nu_i$ is the ionization potential of the level $i$. Care must be taken
in the case of recombination from an excited level into the next lower ion. Then
$(n_i/n_j)^\star=n_e\phi_i \cdot \phi_1^+/\phi_j$.

Dielectronic recombination is included following Mihalas \& Hummer (1973).
Assuming now that $j$ is a ground state of ion $k$, then the recombination rate
into level $i$ of ion $k-1$ via an autoionization level $c$ (with ionization
potential $h\nu_c$, having a negative value when lying above the ionization
limit) is:
\begin{equation}
R_{ji}=\frac{8\pi^2e^2}{mc^3}n_e\phi_if_{ic}e^{h(\nu_c-\nu_i)/kT}
\nu_c^2\left(1+\frac{c^2}{2h\nu_c^3\bar{J}}\right) .
\end{equation}
The reverse process, the autoionization rate, is given by:
\begin{equation}
R_{ij}=\frac{4\pi^2e^2}{hmc}\frac{1}{\nu_c}f_{ic}\bar{J} .
\end{equation}
The oscillator strength for the stabilizing transition (i.e.\ transition
$i\rightarrow c$) is denoted by $f_{ic}$, and $\bar{J}$ is the mean intensity
averaged over the line profile. Our program simply takes $J_{\nu}$ from the
continuum frequency point closest to the transition frequency, which is
reasonable because the autoionization line profiles are extremely broad. The
population of autoionization levels is assumed to be in LTE and therefore such
levels do not appear explicitly in the rate equations.

The computation of collisional rates is generally dependent on the specific ion
or even transition. Several options, covering the most important cases, may be
chosen by the user.

\subsubsection{Abundance-definition equation}

The rate equation for the highest level of a given chemical species is
redundant. It is replaced by the abundance definition equation. 
Summation over all levels usually
includes not only NLTE levels but also levels which are treated in LTE,
according to the specification in the model atom. Denoting the number of
ionization stages of species $k$ with $NION(k)$, the number of NLTE and LTE
levels per ion with $NL(l)$ and $LTE(l)$, respectively, we can write:
\begin{equation}
\sum_{l=1}^{NION(k)}\left[\sum_{i=1}^{NL(l)}n_{kli}+
\sum_{i=1}^{LTE(l)}n_{kli}^{\star}\right]
=y_k (N-n_e) . 
\end{equation}
$y_k$ is the number fraction of element $k$.

\subsubsection{Charge conservation}

We close the system of statistical equilibrium equations by demanding charge
conservation. We denote the total number of chemical species with $\textrm{\it
NATOM}$, the
charge of ion $l$ with $q(l)$ (in units of the electron charge) and write:
\begin{equation}
\sum_{k=1}^{\textrm{\scriptsize\it NATOM}\phantom{)}}\sum_{l=1}^{NION(k)}q(l)
\left[\sum_{i=1}^{NL(l)}n_{kli}+\sum_{i=1}^{LTE(l)}n_{kli}^{\star}\right]=n_e .
\end{equation}

\subsubsection{Complete statistical equilibrium equations}

We introduce a vector comprising the occupation numbers of all NLTE levels,
$\bf n$ $=(n_1, \ldots ,n_{NL})$. Then the statistical equilibrium equation is
written as:
\begin{equation}\label{amat}
\bf An=b .
\end{equation}
The gross structure of the rate matrix $\bf A$ is of block matrix form, because
transitions between levels occur within one ionization stage or to the ground
state of the next ion. The structure is complicated by ionizations into excited
levels and by the abundance definition and charge conservation equations which
give additional non-zero elements in the corresponding lines of ${\bf A}$.

\subsection{Radiative equilibrium}

Radiative equilibrium can be enforced by adjusting the temperature
stratification either during the linearization procedure or in between ALI
iterations. In the former case a linear combination of two different
formulations is used and in the latter case the classical Uns\"old-Lucy
temperature correction procedure (Lucy 1964) is utilized. The latter is
particularly interesting, because it allows one to exploit the block form of the
rate-coefficient matrix. This fact allows an economic block-by-block solution
followed by a subsequent Uns\"old-Lucy temperature correction step. On the
other hand, however, this correction procedure may decelerate the global
convergence behavior of the ALI iteration.

The two forms of expressing the radiative equilibrium condition follow from the
requirement that the energy emitted by a volume element per unit time is equal
to the absorbed energy per unit time (integral form):
\begin{equation}\label{int}
\int_{0}^{\infty}\chi_{\nu}(S_{\nu}-J_{\nu})\,d\nu=0 ,
\end{equation}
where scattering terms in $\chi_\nu$ and $S_\nu$ cancel out. In principle, this
formulation is equivalent to demanding flux-constancy throughout the atmosphere
(differential form)
\begin{equation}\label{diff}
\int_{0}^{\infty}\frac{\partial}{\partial\tau_{\nu}}(f_{\nu}J_{\nu})\,d\nu-{\cal H}=0,
\end{equation}
where ${\cal H}$ (Eq.\,\ref{nominal}) is the nominal flux. Here $f_{\nu}$ is
the variable Eddington factor, defined as
\begin{equation}\label{eddfac}
f_{\nu}=\int_{0}^{1}\mu^2u_{\nu\mu}\,d\mu \Big/ \int_{0}^{1}u_{\nu\mu}\,d\mu,
\end{equation}
which is computed from the Feautrier intensity $u_{\nu\mu}$ (Eq.\,\ref{udef})
after the formal solution. As discussed e.g.~in Hubeny (1988) the differential
form
is more accurate at large depths, while the integral form behaves numerically
superior at small depths. Instead of arbitrarily selecting that depth in the
atmosphere where we switch from one formulation to the other, we use a linear
combination of both constraint equations, which guarantees a smooth transition
with depth, based on physical grounds (Hubeny \& Lanz 1992, Hamann 1994):
\begin{equation}\label{combi}
\frac{1}{\bar\kappa_J}\int_{0}^{\infty}\chi_{\nu}(S_{\nu}-J_{\nu})\,d\nu+
{\bar\Lambda_J^\star}\int_{0}^{\infty}\frac{\partial}{\partial\tau_{\nu}}(f_{\nu}J_{\nu})\,d\nu-
{\bar\Lambda_J^\star}{\cal H}-F_0=0 .
\end{equation}
For details on this equation and on our implementation of the Uns\"old-Lucy
procedure see Dreizler (this volume).

We note that explicit depth coupling is introduced by the differential form
Eq.\,\ref{diff} through the derivative $\partial/\partial\tau_{\nu}$ even if a
purely local $\Lambda^{\star}$ operator is used. Therefore the linearization
procedure can no longer be performed independently at each depth point and the
question at which boundary to start becomes relevant. Numerical experience
shows that it is essential to start at the outer boundary and proceed 
inwards. If a tri-diagonal operator is used, nearest neighbor depth coupling is
introduced anyhow.

\subsection{Hydrostatic equilibrium}\label{defm}

We write the equation for hydrostatic equilibrium as Mihalas (1978, p.\,170):
\begin{equation}
\frac{dP}{dm}=g .
\end{equation}
$P$ is the total
pressure comprising gas, radiation and turbulent pressures, so that:
\begin{equation}\label{hydros}
\frac{d}{dm}\left( NkT+\frac{4\pi}{c}\int_{0}^{\infty}f_{\nu}J_{\nu}\,d\nu
+\frac{1}{2}\rho v^2_{\rm turb}\right)=g
\end{equation}
with Boltzmann's constant $k$ and the turbulent velocity $v_{\rm turb}$. The
hydrostatic equation may either be solved simultaneously with all other
equations or separately between iterations. The overall convergence behavior
is usually the same in both cases. If taken into the linearization scheme, and a
local $\Lambda^{\star}$ operator is used, then, as in the case of the
radiative equilibrium equation, explicit depth-coupling enters via the depth
derivative $d/dm$. Again, solution of the linearized equations must proceed
inwards starting at the outer boundary. The starting value at the first depth
point (subscript $d=1$) is:
\begin{equation}\label{eddfach}
N_1kT_1+\frac{1}{2}\rho_1v^2_{\rm turb}(m_1)=m_1\left(g-
\frac{4\pi}{c}\int_{0}^{\infty}
\frac{\chi_{1,\nu}}{\rho_1}h_{\nu}J_{\nu,k}\,d\nu\right),
\end{equation}
where $h_{\nu}$ is the variable Eddington factor denoting the ratio of
$H_{\nu}/J_{\nu}$ at the surface, kept fixed during linearization.

We are also able to account for element separation resulting from pressure
diffusion, which is the process that governs the spectroscopic appearance
e.g.~of white dwarfs. The numerical method and some results are described by
Dreizler \& Wolff (1999) and Schuh \etal (this volume).

\subsection{Particle-conservation and fictitious massive-particle density}

The total particle density $N$ is the sum of electron density plus the
populations of all atomic/ionic states, in all LTE and NLTE levels: 
\begin{equation}
N=n_e+
\sum_{k=1}^{\textrm{\scriptsize\it NATOM}\phantom{)}}\sum_{l=1}^{NION(k)}
\left[\sum_{i=1}^{NL(l)}n_{kli}+\sum_{i=1}^{LTE(l)}n_{kli}^{\star}\right] .
\end{equation}
The fictitious massive-particle density $n_H$ is introduced for notational
convenience. It is defined by:
\begin{equation}
n_H=\sum_{k=1}^{\textrm{\scriptsize\it NATOM}\phantom{)}}m_k\sum_{l=1}^{NION(k)}
\left[\sum_{i=1}^{NL(l)}n_{kli}+\sum_{i=1}^{LTE(l)}n_{kli}^{\star}\right] .
\end{equation}
The mass of a chemical species in AMU is denoted by $m_k$. Introducing the mass
of a hydrogen atom $m_H$, we may write the material density simply as
\begin{equation}
\rho=n_Hm_H .
\end{equation}

\subsection{Opacity and emissivity \label{opa}}

Thermal opacity and emissivity are made up by radiative bound-bound, bound-free
and free-free transitions. For each species we compute and sum:
\begin{eqnarray}\label{chi}
\kappa_{\nu}&=&\sum_{l=1}^{NION\phantom{)}}\left[
\sum_{i=1}^{NL(l)}\sum_{j>i}^{NL(l)}\sigma_{li\rightarrow lj}(\nu)
\left(n_{li}-n_{lj}\frac{g_{li}}{g_{lj}}
e^{-h(\nu-\nu_{ij})/kT}
\right) \right. \\ 
& & + \sum_{i=1}^{NL(l)}\sum_{j>i}^{NL(l+1)}\sigma_{li\rightarrow l+1,k}(\nu)
\left(n_{li}-n_{li}^{\star}e^{-h\nu/kT}\right) \nonumber \\ 
& & + \left. n_e\sigma_{kk}(l,\nu)\left(1-e^{-h\nu/kT}\right)
\left(\sum_{i=1}^{NL(l+1)}n_{l+1,i}+\sum_{i=1}^{LTE(l+1)}n_{l+1,i}^{\star}
\right)
\right] \nonumber
\end{eqnarray}
where the total opacity includes Thomson scattering, i.e.\
$\chi_\nu=\kappa_\nu+n_e\sigma_e$, and
\begin{eqnarray}\label{eta}
\frac{\eta_{\nu}}{2h\nu^3/c^2}&=&\sum_{l=1}^{NION\phantom{)}}\left[
\sum_{i=1}^{NL(l)}\sum_{j>i}^{NL(l)}\sigma_{li\rightarrow lj}(\nu)
n_{lj}\frac{g_{li}}{g_{lj}}
e^{-h(\nu-\nu_{ij})/kT} 
\right. \\
& & +\sum_{i=1}^{NL(l)}\sum_{j>i}^{NL(l+1)}\sigma_{li\rightarrow l+1,k}(\nu)
n_{li}^{\star}e^{-h\nu/kT} \nonumber \\
& & + \left. 
n_e\sigma_{kk}(l,\nu)e^{-h\nu/kT}
\left(\sum_{i=1}^{NL(l+1)}n_{l+1,i}+\sum_{i=1}^{LTE(l+1)}n_{l+1,i}^{\star}
\right)
\right] . \nonumber
\end{eqnarray}
$\sigma_{li\rightarrow
l+1,k}(\nu)$ denotes the cross-section for photoionization from level $i$ of
ion $l$ into level $k$ of ion $l+1$. The double summation over the bound-free
continua takes into account the possibility that a particular level may be
ionized into more than one level of the next high ion. Again, note the
definition of the LTE population number $n_{li}^{\star}$ in this case, which
depends on the level $(l+1,k)$ of the parent ion:
\begin{equation}
n_{li}^{\star}=n_{l+1,k}n_e\phi_{li}\frac{\phi_{l+1,1}}{\phi_{l+1,k}} .
\end{equation}
Note also that the concept of LTE levels (whose occupation numbers enter,
e.g.~the number- and charge-conservation equations) in the atomic models of
complex ions is therefore not unambiguous. Our code always assumes that
LTE levels in the model atoms are populated in LTE with respect to the ground
state of the upper ion.

The source function used for the approximate radiation transfer is
$\eta_{\nu}/\kappa_{\nu}$, thus it excludes Thomson scattering. For the exact
formal solution of course, the total opacity $\chi_\nu$ in the expression
Eq.\,\ref{source} includes the Thomson term ($n_e\sigma_e$).

\subsection{Atomic level dissolution by plasma perturbations}

As high-lying atomic levels are strongly perturbed by other charged particles
in the plasma they are broadened and finally dissolved. This effect is
observable by line merging at series limits and must be accounted for in line
profile analyses. Moreover, line-overlap couples the radiation field in many
lines and flux-blocking can strongly affect the global atmospheric structure.
Numerically, we treat the level dissolution in terms of occupation
probabilities, which for LTE plasmas can be defined as the ratio of the level
populations to those in absence of perturbations. A phenomenological theory for
these quantities was given in Hummer \& Mihalas (1988). The non-trivial
generalization to NLTE plasmas was made by Hubeny \etal (1994). In practice an
individual occupation-probability factor (depending on $T, n_e$, and principal
quantum number), is applied to each atomic level, which describes the
probability that the level is dissolved. Furthermore, the rate equations
Eq.\,\ref{rates} must be generalized in a unique and unambiguous manner. For
details see Hubeny \etal (1994). 

\section{The Accelerated Lambda Iteration (ALI)}

In all constraint equations described above the mean intensities $J_{\nu}$ are
replaced by the approximate radiation field Eq.\,\ref{ali} in order to eliminate
these variables from the solution vector Eq.\,\ref{psi1}. In principle the ALO
may be of arbitrary form as long as the iteration procedure converges. In
practice, however, an optimum choice is desired in order to achieve
convergence with a minimum amount of iteration steps. The history of the ALOs
is interesting, and was summarized in detail by Hubeny (1992). Of utmost
importance were two papers by Olson and collaborators (Olson \etal 1986, Olson
\& Kunasz 1987) who overcame the major drawback of early ALOs, namely the
occurrence of free parameters controlling the convergence process, and thus
found the optimum choice of ALOs. Our model atmosphere program permits the use
of either a diagonal or a tri-diagonal ALO, both are set up following Olson \&
Kunasz (1987).

\subsection{Diagonal (local) lambda operators}

In this case the mean intensity $J_d$ at a particular depth  $d$ in the current
iteration step is computed solely from the local source function $S_d$ and a
correction term $\Delta J_d$, the latter depending on the source functions (at
all depths) from the previous iteration. Dropping the iteration count and
introducing indices denoting depth points we can rewrite Eq.\,\ref{ali} as:
\begin{equation}
J_d=\Lambda^{\star}_{d,d}S_d+\Delta J_d.
\end{equation}
In discrete form we can think of $\Lambda^{\star}$ as a matrix acting on a
vector whose elements comprise the source functions of all depths. Then
$\Lambda^{\star}_{d,d}$ is the diagonal element of the $\Lambda^{\star}$ matrix
corresponding to depth point $d$. Writing $\Lambda^{\star}_{d,d}\equiv B_d$
(for numerical computation see Eq.\,\ref{matrix} below) we have a purely local
expression for the mean intensity:
\begin{equation}
J_d=B_dS_d+\Delta J_d.
\end{equation}

\subsection{Tridiagonal (non-local) lambda operators}

Much better convergence is obtained if the mean intensity is computed not only
from the local source function, but also from the source function of the
neighboring depths points. Then the matrix representation of $\Lambda^{\star}$
is of tri-diagonal form and we may write:
\begin{equation}\label{trij}
J_d=C_{d-1}S_{d-1}+B_dS_d+A_{d+1}S_{d+1}+\Delta J_d \ ,
\end{equation}
where $C_{d-1}$ and $A_{d+1}$ represent the upper and lower subdiagonal
elements of $\Lambda^{\star}$, and $S_{d\pm 1}$ the source functions at the
adjacent depths. By analogy the correction term becomes:
\begin{equation}\label{correc}
\Delta J_d=\Lambda_{d,d'}S_{d'}-(C_{d-1}S_{d-1}+B_dS_d+A_{d+1}S_{d+1}).
\end{equation}
Again all quantities for the computation of $\Delta J_d$ are known from the previous
iteration, and the first term gives the exact formal solution of the transfer
equation. We emphasize again that the source functions in Eq.\,\ref{trij} are
to be computed from the correct occupation numbers and temperature, all of
which are still unknown. We therefore have a non-linear set of equations
for these quantities, which must be solved by either Newton-Raphson iteration
or other techniques, resulting in the solution of a tri-diagonal linear
equation of the form Eq.\,\ref{tri}.

As was shown in Olson \etal (1986) the elements of the optimum $\Lambda^{\star}$
matrix are given by the corresponding elements of the exact $\Lambda$ matrix.
The diagonal and subdiagonal elements are computed following Olson \& Kunasz
(1987):
\begin{eqnarray}\label{matrix}
A_{d+1}&= &\int_{0}^{1}\left(
e^{-\Delta\tau_d}\frac{e^{-\Delta\tau_{d-1}}-1}{\Delta\tau_{d-1}}
-\frac{e^{-\Delta\tau_d}-1}{\Delta\tau_d} \right)\frac{d\mu}{2}, \nonumber\\
1-B_{d}&=&\int_{0}^{1}\left(
\frac{1-e^{-\Delta\tau_{d-1}}}{\Delta\tau_{d-1}}
+\frac{1-e^{-\Delta\tau_d}}{\Delta\tau_d} \right)\frac{d\mu}{2}, \nonumber\\
C_{d-1}&= &\int_{0}^{1}\left(
e^{-\Delta\tau_{d-1}}\frac{e^{-\Delta\tau_{d}}-1}{\Delta\tau_{d}}
-\frac{e^{-\Delta\tau_{d-1}}-1}{\Delta\tau_{d-1}}\right)\frac{d\mu}{2},     
\end{eqnarray}
with $\Delta\tau_{d-1}\equiv (\tau_d-\tau_{d-1})/\mu$. At large optical depths
with increasing $\Delta\tau$ steps (the depth grid is roughly equidistant in
$\log\tau$) the subdiagonals $A_{d+1}$ and $C_{d-1}$ vanish, and the diagonal
$B_d$ approaches unity, reflecting the fact that the radiation field is more
and more determined by local properties of the matter. At very small optical
depths all elements of $\Lambda^{\star}$ vanish, reflecting the non-localness
of the radiation field at these depths.

\subsection{Acceleration of convergence}

Our code takes advantage of an acceleration scheme to speed up convergence of
the iteration cycle Eq.\,\ref{ali}. We implemented the scheme originally
proposed by Ng (1974) and later by Auer (1987). It extrapolates the correction
vector $\delta\mbox{\boldmath$\psi$}_d$ from the previous three iterations.
From our experience the extrapolation often yields over-corrections, resulting
in alternating convergence or even divergence. In contrast, use of a
tri-diagonal ALO usually results in satisfactorily fast convergence, so that
the acceleration scheme is rarely used.

\section{Solution of the non-linear equations by iteration}

At each depth the complete set of non-linear equations Eq.\,\ref{ali} for a
single iteration step comprises the equations for statistical,
radiative, and hydrostatic equilibrium, and the particle-conservation equation.
For the numerical solution we introduce discrete depth and frequency grids. The
equations are then linearized and solved by a suitable iterative scheme.
Explicit angle-dependence of the radiation field is not required here and is
consequently eliminated by the use of variable Eddington factors. Angle
dependence is considered only in the formal solution of the transfer equation.
Our program requires an input model-atmosphere structure as a starting
approximation, together with an atomic data file, and a frequency grid.
Depth and frequency grids are therefore set up in advance by separate programs.

\subsection{Discretization} 

After a depth-grid is chosen, we start by computing a gray LTE continuum model
using the Uns\"old-Lucy temperature-correction procedure. Depth points (typical
number is 90) are chosen in equidistant steps on a logarithmic
(Rosseland) optical depth scale. The converged LTE model (temperature and
density structure, given on a column mass depth scale) is written to a file
that is read by {\tt PRO2}. The NLTE code uses the column mass as an
independent depth variable.

The frequency grid is determined by the atomic-data input file.
Frequency points are set blue- and red-ward of each absorption edge, and for
each spectral line. Gaps are filled up by setting continuum points. Finally,
the quadrature weights are computed. Frequency integrals appearing e.g.~in
Eq.\,\ref{combi} are replaced by quadrature sums, and depth derivatives
by difference quotients.

\subsection{Linearization}

All variables $x$ are replaced by $x\rightarrow x+\delta x$ where $\delta x$
denotes a small perturbation of $x$. Terms not linear in these perturbations
are neglected. The perturbations are expressed by perturbations of the basic
variables:
\begin{equation}\label{deltax}
\delta x=      \frac{\partial x}{\partial T  }\delta T  +
               \frac{\partial x}{\partial n_e}\delta n_e+
               \frac{\partial x}{\partial N  }\delta N  +
               \frac{\partial x}{\partial n_H}\delta n_H+
\sum_{l=1}^{NL}\frac{\partial x}{\partial n_l}\delta n_l .
\end{equation}
As an illustrative example we linearize the equation for radiative equilibrium.
Most other linearized equations may be found in Werner (1986). Assigning two
indices ($d$ for depth and $i$ for frequency of a grid with NF points) to the
variables and denoting the quadrature weights with $w_i$ Eq.\,\ref{combi}
becomes:
$$
\sum_{i=1}^{NF}
w_i(\frac{\chi{_{di}}}{\bar\kappa_J}[\delta S{_{di}}-\delta J{_{di}}]+\delta\chi{_{di}}[S{_{di}}-J{_{di}}]) 
+{\bar\Lambda_J^\star}\sum_{i=1}^{NF} 
\frac{w_i}{\Delta\tau_i}(\delta J{_{di}} f{_{di}}-\delta J{_{d-1,i}} f{_{d-1,i}}) \nonumber
$$
\begin{equation}\label{lin}
= F_0+{\bar\Lambda_J^\star}{\cal H}- 
\sum_{i=1}^{NF}w_i\frac{\chi{_{di}}}{\bar\kappa_J}(S{_{di}}-J{_{di}})
-{\bar\Lambda_J^\star}\sum_{i=1}^{NF}\frac{w_i}{\Delta\tau_i}(f{_{di}} J{_{di}}-f{_{d-1,i}} J{_{d-1,i}}) .
\end{equation}
Note that we do not linearize $\Delta\tau_i$. Because of this, convergence
properties may deteriorate significantly in some cases. Perturbations
$\delta S{_{di}}, \delta\chi{_{di}}$ are expressed by Eq.\,\ref{deltax}, and the
perturbation of the mean intensity $J{_{di}}$ is, according to Eq.\,\ref{trij},
given through the perturbations of the source function at the current and the
two adjacent depths:
\begin{equation}
\delta J{_{di}}=C{_{d-1,i}}\delta S{_{d-1,i}}+B{_{di}}\delta S{_{di}}+A{_{d+1,i}}\delta S{_{d+1,i}} \ ,
\end{equation}
where $A, B, C$ are the $\Lambda$ matrix elements from Eq.\,\ref{matrix}. The
$\delta J{_{d-1,i}}$ contain the term $C{_{d-2,i}}\delta S{_{d-2,i}}$ which is
neglected because we want to account only for nearest neighbor coupling. We
write $\delta S{_{di}}$ with the help of Eq.\,\ref{deltax} and observe that for
any variable $z$:
\begin{equation}
\frac{\partial S{_{di}}}{\partial z}=\frac{1}{\chi{_{di}}}\left(
\frac{\partial\eta{_{di}}}{\partial z}-S{_{di}}\frac{\partial\chi{_{di}}}{\partial z}
\right).
\end{equation}
Derivatives of opacity and emissivity with respect to temperature, electron and
population densities are computed from analytical expressions (see e.g.~Mihalas
\etal 1975, Werner 1987). We finally get from Eq.\,\ref{lin}:
\begin{eqnarray}
 & & 
\delta T{_{d-1,i}}\left\{
\sum_{i}^{NF}-\frac{w_i}{\bar\kappa_J}
\frac{\partial S{_{d-1,i}}}{\partial T}\chi{_{di}} C{_{d-1,i}} \right.\nonumber
\\
 & &\left. +{\bar\Lambda_J^\star}\sum_{i}^{NF}\frac{w_i}{\Delta\tau_i}
(f{_{di}} C{_{d-1,i}}-f{_{d-1,i}} B{_{d-1,i}})\frac{\partial S{_{d-1,i}}}{\partial T}
\right\}+
\nonumber
\\
 & & 
\delta T_d\left\{
\sum_{i}^{NF}\frac{w_i}{\bar\kappa_J}\left[
\frac{\partial S{_{di}}}{\partial T}\chi{_{di}}(1-B{_{di}})+
\frac{\partial\chi{_{di}}}{\partial T}(S{_{di}}-J{_{di}})\right] \right.\nonumber
\\
 & &\left.+{\bar\Lambda_J^\star}\sum_{i}^{NF}\frac{w_i}{\Delta\tau_i}(f{_{di}} B{_{di}}-f{_{d-1,i}} A{_{di}})
\frac{\partial S{_{di}}}{\partial T}
\right\}+
\nonumber\\
 & & 
\delta T{_{d+1,i}}\left\{
\sum_{i}^{NF}-\frac{w_i}{\bar\kappa_J}
\frac{\partial S{_{d+1,i}}}{\partial T}\chi{_{di}} A{_{d+1,i}} \right.\nonumber
\\
& & \left. +{\bar\Lambda_J^\star}\sum_{i}^{NF}\frac{w_i}{\Delta\tau_i}
(f{_{di}} A{_{d+1,i}}-f{_{d-1,i}} B{_{d+1,i}})\frac{\partial S{_{d+1,i}}}{\partial T}
\right\}+
\nonumber
\\
& & 
\delta n_{e_{d-1,i}}\{\cdots\}
    +\delta n_{e_{d,i}}\{\cdots\}
    +\delta n_{e_{d+1,i}}\{\cdots\}+
\nonumber
\\
& & 
\sum_{l=1}^{NL}\delta n_{l_{d-1,i}}\{\cdots\}
    +\sum_{l=1}^{NL}\delta n_{l_{d,i}}\{\cdots\}
    +\sum_{l=1}^{NL}\delta n_{l_{d+1,i}}\{\cdots\}
\nonumber
\\
 &  & 
= {\rm r.h.s.}
\end{eqnarray}
The curly brackets $\{\cdots\}$ denote terms that are similar to those
multiplied by perturbations of the temperature. Instead of partial derivatives
with respect to $T$, they contain derivatives with respect to $n_e$ and the
populations $n_l$. They all represent coefficients of the matrices
{$\mbox{\boldmath$\alpha, \beta, \gamma$}$} in Eq.\,\ref{tri}.

\subsection{Newton-Raphson iteration}

As described in Sect.\,\ref{eins2} the linearized equations have a tri-diagonal
block-matrix form, see Eq.\,\ref{tri}. Inversion of the grand matrix ($\equiv
{\bf T}$ sized $(NN\cdot ND) \times (NN\cdot ND)$, i.e.~about $10^4\times 10^4$
in typical applications) is performed with a block-Gaussian elimination scheme,
which means that our iteration of the non-linear equations is a
multi-dimensional Newton-Raphson method. The problem is structurally simplified
when explicit-depth coupling is avoided by the use of a local ALO, however, the
numerical effort is not reduced much, because in both cases the main effort
lies with the inversion of matrices sized $NN\times NN$. The Newton-Raphson
iteration involves two numerically expensive steps, first setting up the
Jacobian (comprising {$\mbox{\boldmath$\alpha, \beta, \gamma$}$}) and then
inverting it. Additionally, the matrix inversions in Eq.\,\ref{feau} limit
their size to about $NN=250$ because otherwise numerical accuracy is lost. Two
variants recently introduced in stellar atmosphere calculations are able to
improve both numerical accuracy and, best of all, computational speed.

\subsection{Alternative fast solution techniques for non-linear equations:
Broyden-- and Kantorovich--variants}\label{broy}

Broyden's (1965) variant belongs to the family of quasi-Newton methods; it
was first used in model-atmosphere calculations in Dreizler \& Werner (1991),
Hamann \etal 1991, Koesterke \etal (1992). It avoids the repeated set-up of the
Jacobian by the use of an update formula. In addition, it also gives an
update formula for the {\em inverse} Jacobian. In the case of a local ALO the
solution of the linearized system at any depth is:
\begin{equation}\label{yyy}
\delta\mbox{\boldmath$\psi_k$} = \mbox{\boldmath$\beta$}^{-1}_k \, {\bf c_k} .
\end{equation}
Let $\mbox{\boldmath$\beta$}^{-1}_k$ be the $k$-th iterate of the inverse
Jacobian, then an update can be found from:
\begin{equation}\label{xxx}
\mbox{\boldmath$\beta$}_{k+1}^{-1}=\mbox{\boldmath$\beta$}_k^{-1}
+\frac{({\bf s}_k-\mbox{\boldmath$\beta$}_k^{-1}{\bf y}_k)
{\bf\otimes}({\bf s}_k^T\mbox{\boldmath$\beta$}_k^{-1})}{{\bf s}_k^T
\mbox{\boldmath$\beta$}_k^{-1}{\bf y}_k},
\end{equation}
where ${\bf\otimes}$ denotes the dyadic product, and where we have defined:
\begin{eqnarray}
{\bf s}_k  \equiv & \delta\mbox{\boldmath$\psi$}_k 
& \quad \mbox{solution vector of preceding linearization} , \nonumber\\
{\bf y}_k  \equiv & {\bf c}_{k+1}-{\bf c}_k 
& \quad \mbox{difference of actual and preceding residuum} . \nonumber 
\end{eqnarray}
The convergence rate is super-linear, i.e.~slower than the quadratic rate of
the Newton-Raphson method; but this is more than compensated by the tremendous
speed-up for a single iteration step. It is not always necessary to begin the
iteration with the calculation of an exact Jacobian and its inversion.
Experience shows that in an advanced stage of the overall (ALI-) iteration
Eq.\,\ref{ali} (i.e.\ when corrections become small, of the order 1\%) we can
start the linearization cycle Eq.\,\ref{xxx} by using the inverse Jacobian from
the previous overall iteration. Computational speed-up is extreme in this case,
however, it requires storage of the Jacobians of all depths.

More difficult is its application to the tri-diagonal ALO case. Here we have to
update the grand matrix ${\bf T}$ which, as already mentioned, is of block
tri-diagonal form. We cannot update its inverse, because it is never computed
explicitly. Furthermore we need an update formula that preserves the block
tri-diagonal form which is a prerequisite for its inversion by the Feautrier
scheme Eq.\,\ref{feau}. Such a formula was found by Schubert (1970):
\begin{equation}
{\bf T}_{k+1}={\bf T}_k
+\frac{({\bf y}_k-{\bf T}_k{\bf s}_k)\otimes{\bf\bar s}_k^T}
{{\bf\bar s}_k^T{\bf\bar s}_k}, 
\end{equation}
where ${\bf\bar s}_k\equiv {\bf Z}{\bf s}_k$ with the structure matrix ${\bf Z}$
as defined by:
\[ Z_{ij}=\left\{ \begin{array}{r@{\quad {\rm if}\quad}l}
                   1 & T_{ij}\neq 0 \\ 0 & T_{ij}=0 . 
                  \end{array}  \right.\] 
The vectors ${\bf s}_k$ and ${\bf y}_k$ are defined as above but now they span
quantities over all depth instead of a single depth point. With this formula
we obtain new submatrices {$\mbox{\boldmath$\alpha, \beta, \gamma$}$} and ${\bf
c}$ with which the Feautrier scheme Eq.\,\ref{feau} is solved again. This
procedure saves the computation of derivatives. Another feature realized in our
program also saves the repeated inversion of ${\bf q}\equiv
(\mbox{\boldmath$\beta$}_d+\mbox{\boldmath$\gamma$}_d{\bf D}_{d-1})$ by
updating its inverse with the Broyden formula Eq.\,\ref{xxx}. Similar to the
diagonal ALO case it is also possible to pass starting matrices from one
overall iteration Eq.\,\ref{ali} to the next for the update of $\bf T$ and the
matrix ${\bf q}^{-1}$. In both cases the user specifies two threshold values
for the maximum relative correction in $\delta\mbox{\boldmath$\psi$}$ which
cause the program to switch from Newton-Raphson to Broyden stages 1 and 2.
During stage 1, each new overall cycle Eq.\,\ref{ali} is started with an exact
calculation and inversion of all matrices involved, and in stage 2 these
matrices are passed through each iteration.

Another variant, the Kantorovich (1949) method has been introduced into 
model-atmosphere calculations (Hubeny \& Lanz 1992). It is more simple and
straightforward to implement. This method simply keeps the Jacobian fixed
during the linearization cycle; it is surprisingly stable. In fact it turns
out to be even more stable (i.e.\ it can be utilized in an earlier stage of
iteration) than the Broyden method in the tri-diagonal ALO case. The user of
{\tt PRO2} may choose this variant in two stages in analogy to the Broyden
variant. It was found that in stage 2 it is necessary to update the
Jacobian every 5 to 10 overall iterations in order to prevent divergence.

\section{NLTE metal-line blanketing}                                     

Despite the increase in capacity for solving NLTE model-atmosphere problems 
given by the ALI method combined with pre-conditioning techniques, the
blanketing by millions of lines from the iron-group elements
arising from transitions among some $10^5$ levels can be attacked only with
statistical methods. These were introduced into NLTE model-atmosphere
work by Anderson (1989, 1991). At the outset, model atoms are constructed by
combining many thousands of levels into a relatively small number of
{\it superlevels} which can be treated by ALI (or other) methods. Then, in order
to reduce the computational effort, two approaches were developed which vastly
decrease the number of frequency points (hence the number of transfer equations
to be solved) needed to describe properly the complex frequency dependence of
the opacity. These two approaches have their roots in LTE modeling techniques,
where statistical methods are applied for the same reason in the treatment of
opacity: The Opacity Distribution Function (ODF) and Opacity Sampling (OS)
approaches. Both are based on the fact that the opacity (in the LTE
approximation) is a function of two only local thermodynamic quantities.
Roughly speaking, each opacity source can be written in terms of a population
density and a photon cross-section for the respective radiative transition:
{$\kappa_\nu \sim n_l \sigma_{lu}(\nu)$}. In LTE the population follows from
the Saha-Boltzmann equations, hence $n_l=n_l(n_e,T)$. The OS and ODF methods
use such pre-tabulated (on a very fine frequency mesh) $\kappa_\nu(n_e,T)$
during the model atmosphere calculations. The NLTE situation is more
complicated, because pre-tabulation of opacities is not useful. The occupation
numbers at any depth now also depend explicitly on the radiation field (via
the rate equations which replace the TE Saha-Boltzmann statistics) and thus
on the populations in every other depth of the atmosphere. As a consequence, the
OS and ODF methods are not applied to opacity tabulations, but to tabulations
of the photon cross-sections $\sigma(\nu)$. These do depend on local quantities
only, e.g.~line broadening by Stark- and Doppler-effects is calculated from $T$
and $n_e$. In the NLTE case cross-sections take over the role which the
opacity played in the LTE case. So, strictly speaking, the designation OS and
ODF is not quite correct in the NLTE context.

The strategy in our code is the following. Before any model atmosphere
calculation is started, the atomic data are prepared by constructing
superlevels,  and the cross-sections for superlines. Then these cross-sections
are either sampled  on a coarse frequency grid or ODFs are constructed. These
data are put into the model atom which is read by the code. The code does not
know if OS or ODFs are used, i.e.~it is written to be independent of these 
approaches.

Further details about our implementation are described in Dreizler \& Wer\-ner
(1993), Haas \etal (1996), Werner \& Dreizler (1999), and by Rauch \& Deetjen
in this volume.

\section{Summary}

We have presented in detail our technique for numerical solution of the
classical model-atmosphere problem. The construction of metal-line-blanketed
models in hydrostatic and radiative equilibrium under NLTE conditions was the
last and longest-standing problem of classical model atmosphere theory. It 
has finally been solved with a high degree of sophistication. The essential
milestones for this development, starting from the pioneering work of Auer 
\& Mihalas (1969) are:
\begin{itemize}
\item Introduction of Accelerated Lambda Iteration (ALI, or ``operator
splitting'') methods, based upon early work by Cannon (1973) and Scharmer
(1981). The first ALI model atmospheres were constructed by Werner (1986).
\item Introduction of statistical approaches to treat the iron-group elements
in NLTE by Anderson (1989, 1991).
\item Linear formulation of the statistical-equilibrium equations
(Rybicki \& Hummer 1991, Hauschildt 1993).
\item Computation of atomic data by Kurucz (1991), by the Opacity Project
(Seaton \etal 1994) and subsequent improvements, and by the Iron Project (Hummer
\etal 1993).
\end{itemize}

\end{document}